\documentclass[prx,floatfix,showpacs,superscriptaddress,longbibliography,twocolumn]{revtex4-1}

\usepackage[USenglish]{babel}

\usepackage{amssymb, amsmath, amsfonts, amsthm, mathtools, nicefrac, bm, dsfont, txfonts}
\usepackage{tensor}
\usepackage{fixmath}
\usepackage{braket,physics}
\usepackage{siunitx}

\usepackage{epsfig}
\usepackage{subfigure}

\usepackage[colorlinks=true,linkcolor=myblue,citecolor=myred]{hyperref}

\usepackage{float}
\usepackage{appendix}
\newcommand{\matn}{\mathrm}
\newcommand{\opr}{\mathcal}

\renewcommand{\vec}{\boldsymbol}

\usepackage{chngcntr}
\counterwithout{figure}{section}
\usepackage{xcolor}
    \definecolor{myred}  {HTML}{A03658}
    \definecolor{myblue} {HTML} {79A1CD}
    \definecolor{myyellow} {cmy} {0,0.263,0.741}
    \definecolor{mygreen} {HTML}{297C68}
    \colorlet{myorange} {myyellow!60!myred}
    \colorlet{myviolett} {myred!50!myblue!80}
    
\makeatother
    \hypersetup{
      linkcolor = mygreen,
      citecolor  = myred,
      urlcolor   = myred,
      colorlinks = true,
    }

\newcommand{\equ}[1]
{Eq.~(\ref{#1})}

\newcommand{\figu}[1]
{Fig.~\ref{#1}}

\newcommand{\secu}[1]
{Sec.~\ref{#1}}

\newcommand{\be}{\begin{equation}}
\newcommand{\ee}{\end{equation}}

\begin{document}

\date{\today}

\author{Aaron~M\"uller}
\affiliation{Department of Physics, University of Erlangen-N\"urnberg,
  91058 Erlangen, Germany}

\author{Francesco~Grandi}
\affiliation{Department of Physics, University of Erlangen-N\"urnberg,
  91058 Erlangen, Germany}

\author{Martin~Eckstein}
\affiliation{Department of Physics, University of Erlangen-N\"urnberg,
  91058 Erlangen, Germany}

\title{Ultrafast control of spin-orbital separation probed with time-resolved RIXS}

\begin{abstract}
Quasi-one-dimensional systems exhibit many-body effects elusive in higher dimensions. A prime example is spin-orbital separation, which has been measured by resonant inelastic X-ray scattering (RIXS) in Sr\textsubscript{2}CuO\textsubscript{3}. Here, we theoretically analyze the time-resolved RIXS spectrum of Sr\textsubscript{2}CuO\textsubscript{3} under the action of a time-dependent electric field. We show that the external field can reversibly modify the parameters in the effective $t-J$ model used to describe spinon and orbiton dynamics in the material. For strong driving amplitudes, we find that the spectrum changes qualitatively as a result of reversing the relative spinon to orbiton velocity. The analysis shows that in general, the spin-orbital dynamics in Mott insulators in combination with time-resolved RIXS should provide a suitable platform to explore the reversible control of many-body physics in the solid with strong laser fields. 
\end{abstract}
\maketitle

%%%%%%%%%%%%%%%%%%%%%%%%%%%%% 
\textit{Introduction -- }
%%%%%%%%%%%%%%%%%%%%%%%%%%%%% 
Electrons in strongly correlated materials  can be characterized by charge, spin and orbital quantum numbers. In a prototypical Mott insulator, the charge becomes localized by the electron-electron repulsion,  leading to highly intertwined spin-orbital physics and a variety of ordering phenomena \cite{Tokura2000_Science}. In low-dimensional materials, the effect of the Coulomb repulsion can be even more exotic, and the electron can fractionalize into spin and orbital excitations (spin-orbital separation, SOS) \cite{Wohlfeld2011,Schlappa2012} in analogy to spin-charge separation (SCS) \cite{Lieb1968_PRL, Kim1996,Kim2006_NatPhys}. Ultra-short laser pulses have opened avenues to control the many-particle physics in solids out of equilibrium \cite{review_giannetti,Basov2017,delaTorre2021_RMP}. A particularly versatile concept is Floquet engineering, where the Hamiltonian of a system is dynamically modified with time-periodic fields, as employed widely in optical lattice experiments \cite{Jotzu2014_Nat,Georg2019_NatPhys,Struck2012_PRL,Gorg2018}. For example, a control of SCS has been proposed through Floquet engineering of the $t-J$ model \cite{Gao2020_PRL}. In solids, however, a major challenge to implement a similar dynamic control is the heating due to photon absorption from the drive. Few theoretical proposals have so far been realized, such as Floquet Bloch states of weakly interacting electrons \cite{Oka2009_PRB,Wang2013_Science,McIver2020_NatPhys}. Only very recently, a seminal experiment has shown the Floquet control of nonlinear local properties in a strongly correlated compound \cite{Shan2021_Nature}.

In this letter, we show that the spin and orbital degrees of freedom in Mott insulators provide a promising platform to realize a dynamic control of dispersive many-body physics in solids, and in particular, electron fractionalization. As discussed in \cite{Shan2021_Nature}, the Mott gap implies a large transparency window that limits heating via multi-photon absorption, as needed for Floquet engineering. A large gap can also make the insulator robust against tunnelling breakdown in static fields \cite{Oka2010_PRB,Eckstein2010_PRL}. This allows for alternative, less explored pathways to control the low-energy spin-orbital physics with slowly varying strong field transients. Moreover, spin-orbital excitations can be probed using resonant inelastic X-ray scattering (RIXS) \cite{Ament2011,Mitrano2020}, and recent upgrades of free-electron lasers should provide a sufficient time ($\sim 30$ fs) and energy ($\sim 0.05$ eV) resolution to resolve their dynamics \cite{Dunne2019, Dunne2018}.  

We particularly focus on the charge transfer insulator Sr\textsubscript{2}CuO\textsubscript{3}. The spin-orbital excitations in this material are well described by an effective $t-J$ model, where $t$ refers to the hopping matrix element for an orbital excitation, which is very different from the original electronic tunneling, and J is the usual spin exchange constant. The model displays signatures of SOS, as clearly visible in static RIXS \cite{Schlappa2012}. The orbiton in Sr\textsubscript{2}CuO\textsubscript{3} turns out to be slower than the spinon, with a speed comparable in magnitude ($J/t=2.8$). Notice that this situation is very different from the typical case of SCS, where $t$ refers to the original hopping of a charge excitation, which is typically much faster than the spinon. In the case of Sr\textsubscript{2}CuO\textsubscript{3}, the ratio $t/J$ should therefore be controllable over a wide range, eventually allowing for a nontrivial dynamic switch between distinct regimes in which the spinon is either faster or slower than the orbiton.

%%%%%%%%%%%%%%%%%%%%%%%%%%%%%%
\textit{Equilibrium Hamiltonian -- }
%%%%%%%%%%%%%%%%%%%%%%%%%%%%%%
The quasi-one-dimensional charge-transfer insulator Sr\textsubscript{2}CuO\textsubscript{3} can be described as a chain of alternating copper and oxygen atoms in $3$d$^9$ configuration and with filled $2$p orbitals. This makes it convenient to work in a hole representation, with one hole per unit cell. The low point-group symmetry of the crystal entirely lifts the degeneracy of the Cu 3d orbitals, and the ground state is a Mott insulator with one hole in the $3 \matn d_{x^2-y^2}$ orbital (termed $a$ in the following). Without any orbital excitation the  Hamiltonian of the system is a Heisenberg model with antiferromagnetic exchange $J$. The motion of orbital excitations on top of this can be described by an effective $t-J$ model \cite{Wohlfeld2013}, which is found well in agreement with experiment \cite{Schlappa2012},
\be \label{eq:H_t-J}
    \begin{split}
        \opr H_{t-J} = & -t \sum_{j, \sigma} \left( p_{j \sigma}^\dagger p_{j+1 \sigma} + h.c. \right)  - E \sum_j  \tilde n_j + \opr H_{s}.
    \end{split}
\ee
Here the fermion operator $p_{j \sigma}$ creates an orbital excitation where a hole with spin $\sigma$ at site $j$ is transferred to another orbital $b$, and $\tilde n_j$ counts the holes in the $a$ orbital; $\vec S_j$ is the spin of the Cu hole, and $\opr H_{s} = J \sum_j  \left( \vec S_j \vec S_{j+1} - \frac{1}{4} \tilde n_j \tilde n_{j+1}\right)$ a Heisenberg Hamiltonian. An analogous Hamiltonian can be written for different orbitons. Here we focus on $b \equiv 3 \matn d_{zx}$, which turns out to have the broadest dispersion among the $3 \matn d$ orbitals. To derive this model \cite{Wohlfeld2013}, the valence bands of Sr\textsubscript{2}CuO\textsubscript{3} are described by an ab initio charge-transfer Hamiltonian $\opr H_{c-t}$ with nearest-neighbor hopping, local Hubbard interaction and Hund's coupling, charge transfer energy between the Cu and oxygen states, and a nearest-neighbor Coulomb repulsion \cite{Neudert2000} (for details, see \cite{Supplementary}). The large electron-electron repulsion justifies a strong coupling expansion, which projects out doubly occupied orbitals and Cu-O charge transfer excitations, and results in a Kugel-Khomskii Hamiltonian \cite{Kugel1982} for the spin and orbital degrees of freedom of the Cu hole. Spin and orbital superexchange proceeds via the O $2$p orbitals and is therefore obtained from a fourth-order perturbation expansion in the p-d hopping \cite{PhysRevB.37.9423}. By neglecting spin flip processes in the hopping of the hole in the $b$ orbital, and choosing a Jordan Wigner representation for the orbital pseudo-spin, the Kugel-Khomskii Hamiltonian is then mapped on the $t-J$ model \equ{eq:H_t-J}. In this way, $J>0$ and $t$ in \equ{eq:H_t-J} relate to the spin and orbital superexchange  interaction, and $E$ is related to the crystal field splitting. When the orbiton is present, the first part of \equ{eq:H_t-J} describes its motion in a Ne\'el antiferromagnetic background. The equilibrium parameters in \equ{eq:H_t-J} for Sr\textsubscript{2}CuO\textsubscript{3} are \cite{Wohlfeld2013} $t \sim 0.085$ eV, $J \sim 0.238$ eV and $E \sim 1.999$ eV. The large ratio $J/t \sim 2.8$  is indeed aberrant compared to the typical value of the spin-charge $t-J$ model obtained from a single band Hubbard model at large interaction $U$ \cite{Chao_1977}. In the latter, the parameter $t$ is the original electron hopping, and $J=4t^2/U \ll t$, 
while in \equ{eq:H_t-J} both $t$ and $J$ originate from a fourth-order perturbation expansion. 
\begin{figure}
    \centering
    \includegraphics[scale=0.158]{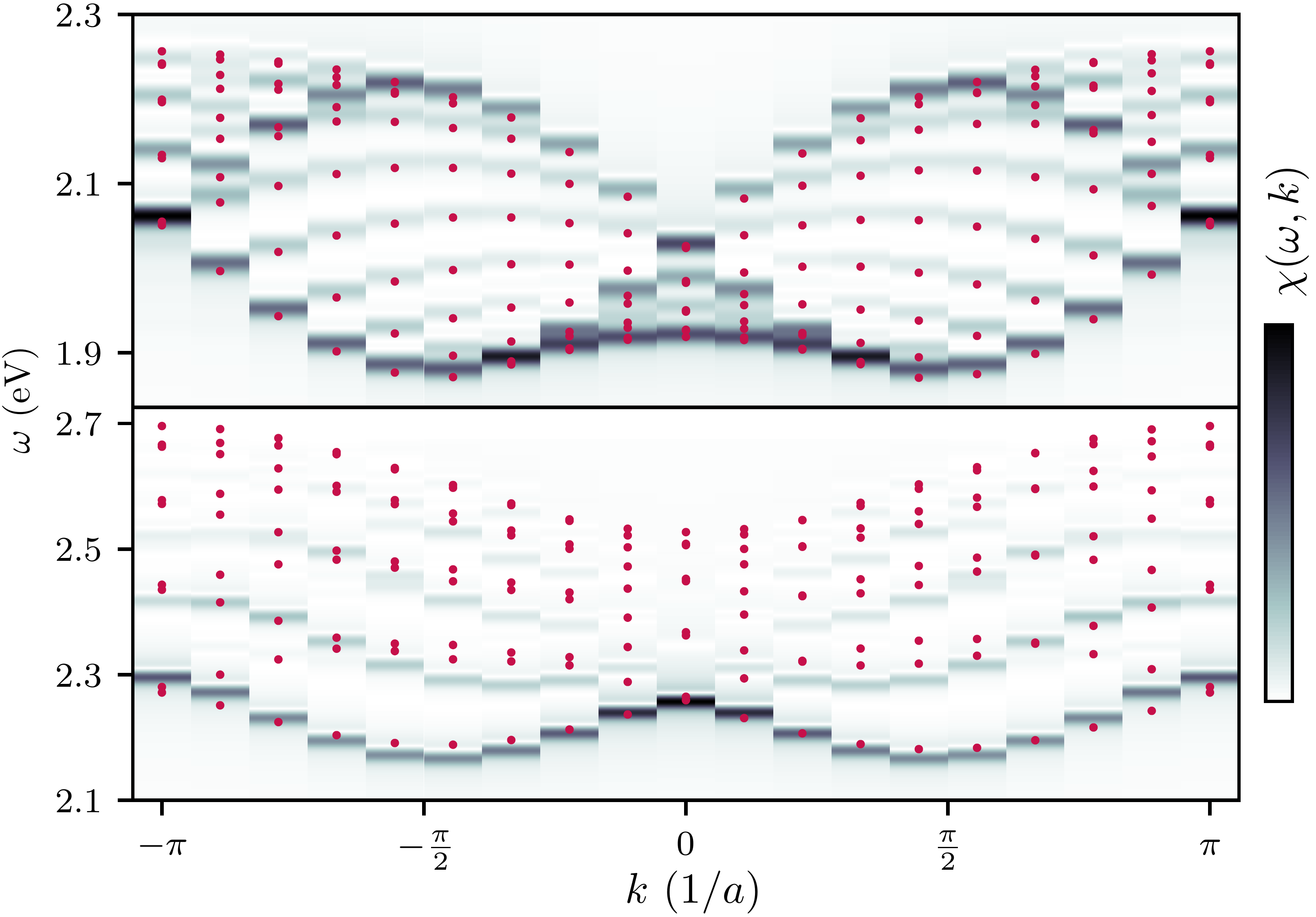}
    \vspace*{-6mm}
    \caption{Spectral function $\chi(\omega, k)$ in the a) slow ($J/t = 0.4$) and b) fast ($J/t = 2.8$) spinon regime, computed on a chain with $L=18$ sites and Gaussian broadening. Red dots are obtained with the spin-orbital separation ansatz with $E_f = E + J$ and a) $t_f = t$, $J_f=J$ and b) $t_f=t/2$, $J_f=J$.} 
    \label{fig:eq_comp_slow_fast}
\end{figure}

Below we will analyze how external fields can be used to control $t$ and $J$ over a wide range, covering both $J/t > 2$ (fast spinon regime) and $J/t < 2$ (slow spinon regime). 

To prepare for this analysis, we first discuss the equilibrium orbital spectral function $\chi(\omega,k) = - 1/\pi \, \text{Im} \, G_{k}(\omega+i0^+)$ of the momentum-dependent single orbiton propagator $G_k$; $G_k$ is the Fourier transform $G_{k} =  1/L \sum_{j, j'= 0}^{L-1} e^{i k (j'-j)} G_{j',j}$ of the real-space propagator
\begin{align} \label{eq:hole_propagator}
    G_{j',j} (t',t) =  -i \Bra{\Psi_0} 
    \opr U_{s}(t_0,t')  p_{j' \sigma}^\dagger \opr U_{t-J}(t',t)p_{j \sigma} \opr U_{s}(t,t_0) \Ket{\Psi_0}.
\end{align}
%In \equ{eq:hole_propagator}, 
Here, $\Ket{\Psi_0}$ is the initial ground state (no orbital excitation) at time $t_0\to-\infty$, and $\opr U_{s}$ and $\opr U_{t-J}$ denote time-evolution in the Heisenberg and $t-J$ model, respectively. The resulting spectrum shows a distinct form in the slow (\figu{fig:eq_comp_slow_fast}a) and fast (\figu{fig:eq_comp_slow_fast}b) spinon regime. 
Note that the spectral function $\chi (\omega,k)$ of the effective $t-J$ model corresponds to the dynamical momentum dependent orbital structure factor of the electronic model. This quantity is entirely unrelated to the electronic spectral function, which measures final states at a different electron number (such as electron removal for an ARPES measurement \cite{Wang2018}).

As a consequence of the spin-orbital separation, insight into the form of $\chi(\omega,k)$ can be acquired from a semi-phenomenological spectral building principle \cite{Eder1997, Bannister2000}: Assuming that the orbital excitation separates into a spinon and an orbiton, which are treated as independent particles with dispersion $\epsilon_s(k_s)=\pi J_f/2 \, \abs{\cos{k_s}}$ for $-\pi/2 \leq k_s \leq \pi/2$ \cite{Giamarchi2004} and $\epsilon_o(k_o)=E_f-2 t_f \, \cos{k_o}$, respectively, spectral weight in  $\chi(\omega,k)$ should be found at energies $\epsilon(k) = \epsilon_s(k_s) + \epsilon_o(k_o)$, with a constraint $k=k_s+k_o$ due to momentum conservation. Hence, one expects the peaks of $\chi(\omega,k)$ at $\epsilon(k)=E_f - 2 t_f \cos{k_o} + J_f \frac{\pi}{2} \abs{\cos{(k-k_o)}}$. A $1/L$-correction to the momentum of the orbiton has to be applied on account of periodic boundary conditions \cite{Giamarchi2004, Bohrdt2018}. In \figu{fig:eq_comp_slow_fast}, $\epsilon(k)$ is shown by red dots. In the slow spinon regime, the spectrum is enclosed by the lower and upper orbiton branches \cite{Suzuura1997, Kim2006_NatPhys}, and there is an overall good agreement between the numerical results and $\epsilon(k)$. In the fast spinon regime, the spectral building principle still reproduces the lower bound of the spectrum, although the parameter $t_f$ needs to be renormalized with respect to the bare $t$ due to further dressing of the orbiton \cite{Brunner2000}. The transition between these two regimes can be located at the so-called supersymmetric point $J/t = 2$, where the $t-J$ model becomes exactly solvable by Bethe Ansatz \cite{Bares1991}.

\begin{figure} 
    \centering
    \includegraphics[scale=0.127]{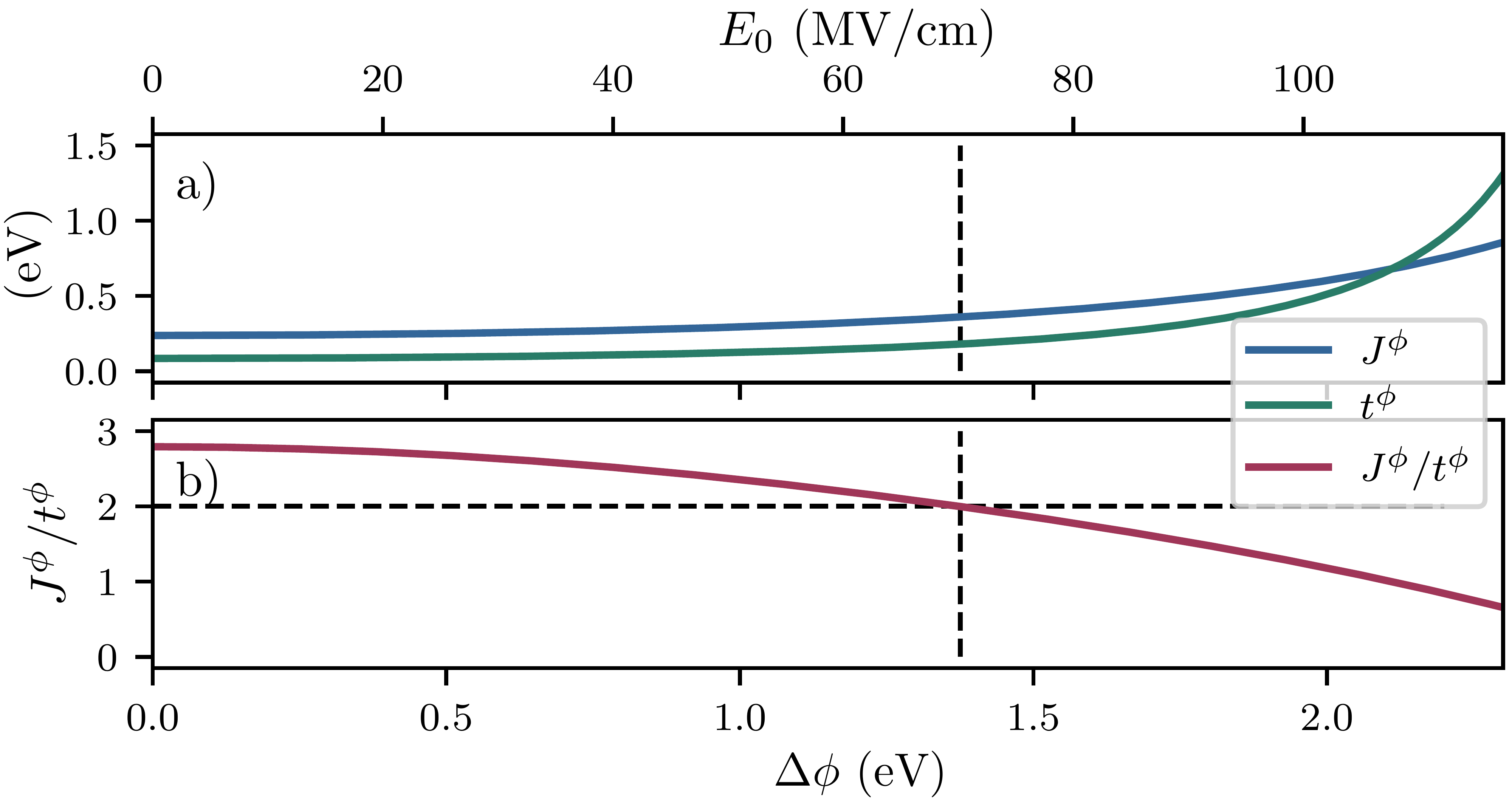}
    \vspace*{-6mm}
    \caption{a) Effective parameters $J^\phi$ and $t^\phi$ versus the potential field gradient $\Delta \phi$ (electric field amplitude $E_0$) in the lower (upper) horizontal axis for the quasi-static regime and b)  the corresponding change in the characteristic value $J/t$. The horizontal dashed line shows the value  $J/t=2$ separating slow and fast spinon regime.}
    \label{fig:qs_micParam}
\end{figure}

\textit{Non-equilibrium effective Hamiltonian --} To include the effect of the driving laser, we couple the charge transfer Hamiltonian $\opr H_{c-t}$ to an electromagnetic field using the Peierls substitution \cite{Peierls1933, Li2020_Coup}. In the length gauge within the electric-dipole approximation \cite{PhysRevA.22.1786,Boykin2001_PRB,baykusheva2020attosecond}, this is equivalent to adding a scalar potential $\Phi_j(t)=- e E(t) X_j$ at each site (independent of the orbital), where $e$ is the elementary charge, $E(t)$ the time-dependent electric field, and  $X_j = j \tilde a /2$ represents the position along the chain, with the copper-copper distance $\tilde a \sim 0.392$ nm \cite{PhysRevB.56.3402}. We neglect the renormalization of the hopping due to dipole matrix elements. Below, we parametrize the time-dependence of the electric field as $E(t)=S(t) \cos( \Omega t)$, with an oscillatory part and an envelope $S(t)$ with maximum amplitude $E_0$. The time-dependent modification of the parameters $t$ and $J$ is then understood in two distinct limits:

In the quasi-static limit, one can derive a $t-J$ Hamiltonian at a given time $t$ by repeating the strong coupling perturbation expansion including energy shifts of the orbitals caused by the instantaneous scalar potential difference $\Delta \phi = e \tilde a E(t) / 2$ between neighboring sites. The potential modifies the virtual intermediate state of the perturbation expansion, and yields field-dependent parameters $t^\phi$, $J^\phi$ and $E^\phi$ in \equ{eq:H_t-J}. (For explicit expressions, see the Supplemental Material \cite{Supplementary}.) 
This quasi-static argument should  apply to slowly varying fields $\Omega \ll t,J$, up to several $10$ THz given the values of $t$ and $J$. Both $J^\phi$ and $t^\phi$ increase with $\Delta \phi$, leading to an overall decrease of the ratio $J/t$ (\figu{fig:qs_micParam}). The intrinsic limitation to the quasi-static control of the spin-orbital physics is the tunnelling breakdown of the insulating state, which would lead to a  population of real charge transfer excitations. The most rapid breakdown is expected when the gradient $\Delta \phi$ is resonant to the nearest-neighbour charge transfer energy $E_{CT}$. Simulations in one- and two-band Hubbard models  consistently show that the carrier generation rate becomes exponentially suppressed for fields sufficiently below $E_{CT}$ \cite{Oka2010_PRB,Eckstein2010_PRL,Dasari2020_PRB}. The fields needed to decrease the ratio $J/t$ below the supersymmetric point are well below this resonant amplitude $E_{CT} \sim 130$ MV/cm ($\Delta \phi_{CT} = 2.6$ eV) for the present system, so that a control of the spin-orbital physics should be possible at least transiently.

%%%%%%%%%%%%%%%%%%%%%%%%%%%%% Floquet
In the Floquet limit, one can conveniently write a time-periodic Hamiltonian in the Floquet basis by moving from the Fock space $\mathcal{F}$ of the original problem to a larger space $\mathcal{F} \otimes \mathcal{T}$, with $\mathcal{T}$ the Hilbert space of the square-integrable periodic functions with period $T = 2 \pi / \Omega$. States in the extended space are labelled by an additional discrete index $n$, which can be understood as the number of photons absorbed or emitted relative to the drive. The representation of $\opr H(t)$ on the enlarged Fock space is called Floquet Hamiltonian $\opr H^{Fl}_{c-t}$, which defines an effective static problem that describes the stroboscopic evolution of the driven system. Since $\opr H^{Fl}_{c-t}$ is time-independent, one can use a strong coupling perturbation expansion analogous to the undriven case to derive the fourth-order spin-orbital superexchange interactions underlying the $t-J$ model \equ{eq:H_t-J}. The Floquet control of superexchange interactions has been derived for various situations, including exchange interactions in the magnetic \cite{Mentink2015, Itin2015_PRL, Bukov2015_AdvPhys} and charge sector \cite{Kitamura2016} of the single-band Hubbard model, antisymmetric exchange interactions which may stabilize chiral spin liquids \cite{Claassen2017_NatComm}, and superexchange via ligand ions \cite{Chaudhary2020}. To derive the superexchange interaction in the Floquet representation in the present case, we project out both virtual charge excitations and states in Floquet sectors $n \neq 0$ (virtual photons). The emission or absorption of $n$ virtual photons during an electronic hopping process shifts the energy of the intermediate states of the superexchange process by $\pm n\Omega$. The matrix elements for such processes are controlled by the Floquet parameter $\mathcal{E} = e \tilde a E_0/ (2 \Omega)$. Explicit calculations and results for the dependence of $t^{\mathcal{E}, \Omega}$, $J^{\mathcal{E}, \Omega}$ and $E^{\mathcal{E}, \Omega}$ on $\mathcal{E}$ and $\Omega$ are provided in the Supplemental Material \cite{Supplementary}. 

\begin{figure}
    \centering
    \includegraphics[scale=0.175]{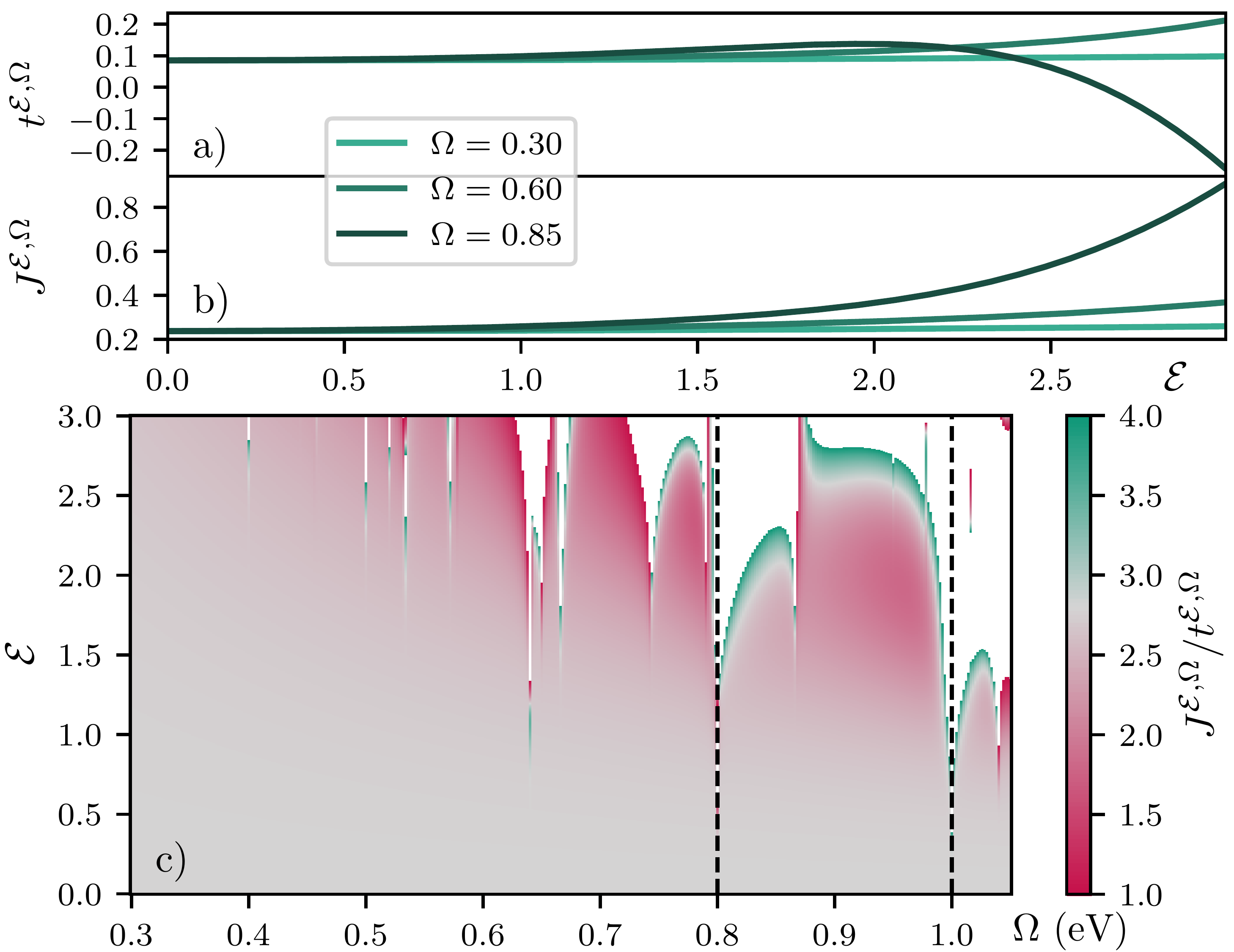}
    \vspace*{-6mm}
    \caption{a), b) Effective parameters $t^{\mathcal{E}, \Omega}$ and $J^{\mathcal{E}, \Omega}$ versus the Floquet parameter $\mathcal{E}$ for several driving frequencies $\Omega$. c) $J^{\mathcal{E}, \Omega}/t^{\mathcal{E}, \Omega}$ as function of $\Omega$ and $\mathcal{E}$. The white areas correspond to regimes in which $J^{\mathcal{E}, \Omega}/t^{\mathcal{E}, \Omega}$ lies outside the indicated range. The dashed lines show the leading Floquet resonances in the given frequency range, which involve a virtual transition to the $n=4$ Floquet sector.}
    \label{fig:flo_micParam}
\end{figure}

In the following, Floquet theory is applied for frequencies $\Omega > t, J $, but below the optical gap $\Delta_{gap} \sim 1.6-1.8$ eV \cite{Imada1998_RMP} to limit charge transfer excitations due to linear photon absorption. By avoiding the Floquet resonances (resonant frequencies for which the intermediate state energy in the superexchange process would vanish), one can further also limit multi-photon absorption. In contrast to what is observed in the quasi-static limit, it is possible to both decrease and increase the hopping $t^{\mathcal{E}, \Omega}$ with respect to the equilibrium value, see \figu{fig:flo_micParam}a), while $J^{\mathcal{E}, \Omega}$ can only increase in the analyzed domain, see \figu{fig:flo_micParam}b). The ratio $J/t$ can therefore be controlled in a wide range below the supersymmetric point $J/t = 2$, see \figu{fig:flo_micParam}c). For example, for $\Omega \sim 0.78$ eV the regime $J/t < 2$ is reached for a Floquet parameter $\mathcal{E} \sim 2$. Divergences of $J/t$ appear upon increasing $\mathcal{E}$ at fixed $\Omega$ due to a localization of the orbiton ($t=0$) when the spin excitations remain mobile \cite{Gao2020_PRL}.
%

%%%%%%%%%%%%%%%%%%%%%%%%%%%%%% Spectral function
%
\textit{Time-resolved RIXS -- }
To verify the dynamic control over the superexchange interactions described in the previous section, one needs an experimental probe able to follow the transient change of the spectrum over the whole Brillouin zone. Time-resolved RIXS (trRIXS) seems particularly suitable for this task. The incoming photon with energy $\omega_i$, momentum $k_i$ and polarization $\vec e_i$ excites the hole originally in the $a$ orbital to the $2$p core state of a copper atom. This highly unstable intermediate state can decay within few femtoseconds to the $b$ orbital by emitting a photon with energy $\omega_f$, momentum $k_f$ and polarization $\vec e_f$, leaving the system with a d-d excitation. In principle, the hole can also decay into other d orbitals, but we are only interested in the $b$ channel, as motivated previously. Thus, measuring the energy loss ($\omega = \omega_i - \omega_f$) and momentum loss ($k = k_i - k_f$) of the photon allows to probe the excitation spectrum of the compound. For the short-lived intermediate state, one can apply the ultra-short core-hole approximation (UCA) \cite{Ament_PRB_2007}, which describes the photon scattering via the intermediate state in terms of a single RIXS operator $\opr B$. The UCA is well established in interpreting static RIXS probes of the dynamics of spin and orbital excitations in equilibrium \cite{Schlappa2012}. With core hole lifetimes in the range of few femtoseconds, the UCA remains justified also for trRIXS of processes on the $100$fs timescale, as in the present case. In the Supplemental Material \cite{Supplementary}, we derive the trRIXS signal within the UCA, starting from the general expression for trRIXS in terms of a four-point correlation function \cite{Chen2019,PhysRevB.103.115136}. The result can be written as $I_{RIXS} = 2 ( \vert B^{\uparrow \uparrow}_{\vec e_i \vec e_f} \vert^2 + \vert B^{\uparrow \downarrow}_{\vec e_i \vec e_f} \vert^2 ) \chi(\omega, k, \bar{t})$. Here $B^{\uparrow \uparrow}_{\vec e_i \vec e_f}$ and $B^{\uparrow \downarrow}_{\vec e_i \vec e_f}$ are the matrix elements of the RIXS operator \cite{Marra2012_PRL,Marra2016}, which will be the same for time-resolved and static RIXS, and $\chi(\omega, k, \bar{t})$ is the convolution of the propagator \eqref{eq:hole_propagator} with probe functions centered around a probe time $\bar t $, $\chi(\omega, k, \bar{t}) =  i \int d t' d t \  s(t', \bar{t}) s(t, \bar{t})  e^{-i \omega (t-t')} G_{k} (t',t)$. The probe-pulse is thereby described semi-classically by the Gaussian envelope $s(t, \bar{t}) = \exp \left(-(t - \bar{t})^2/(2 \sigma_{pr}^2) \right)$.

In the following, instead of analyzing $I_{RIXS}$, we focus on $\chi(\omega, k, \bar{t})$, which does not depend on the details of the experimental apparatus used to measure the RIXS signal. Despite the similarity of the above relation for $I_{RIXS}$ to the angle resolved photo emission spectroscopy signal, the two quantities actually rely on different expressions for the effective hopping parameter $t$, see the Supplemental Material [26]. The numerical evaluation of the single hole propagator \equ{eq:hole_propagator} is done using a Krylov time propagation scheme with middle point approximation for the time evolution \cite{Manmana2005, Alvermann2012, Weinberg2019}.
 
%%%%%%%%%%%%%%%%%%%%%%%%%%%%%% DYNAMIC CONTROL
%
\begin{figure}
    \centering
    \includegraphics[scale=0.177]{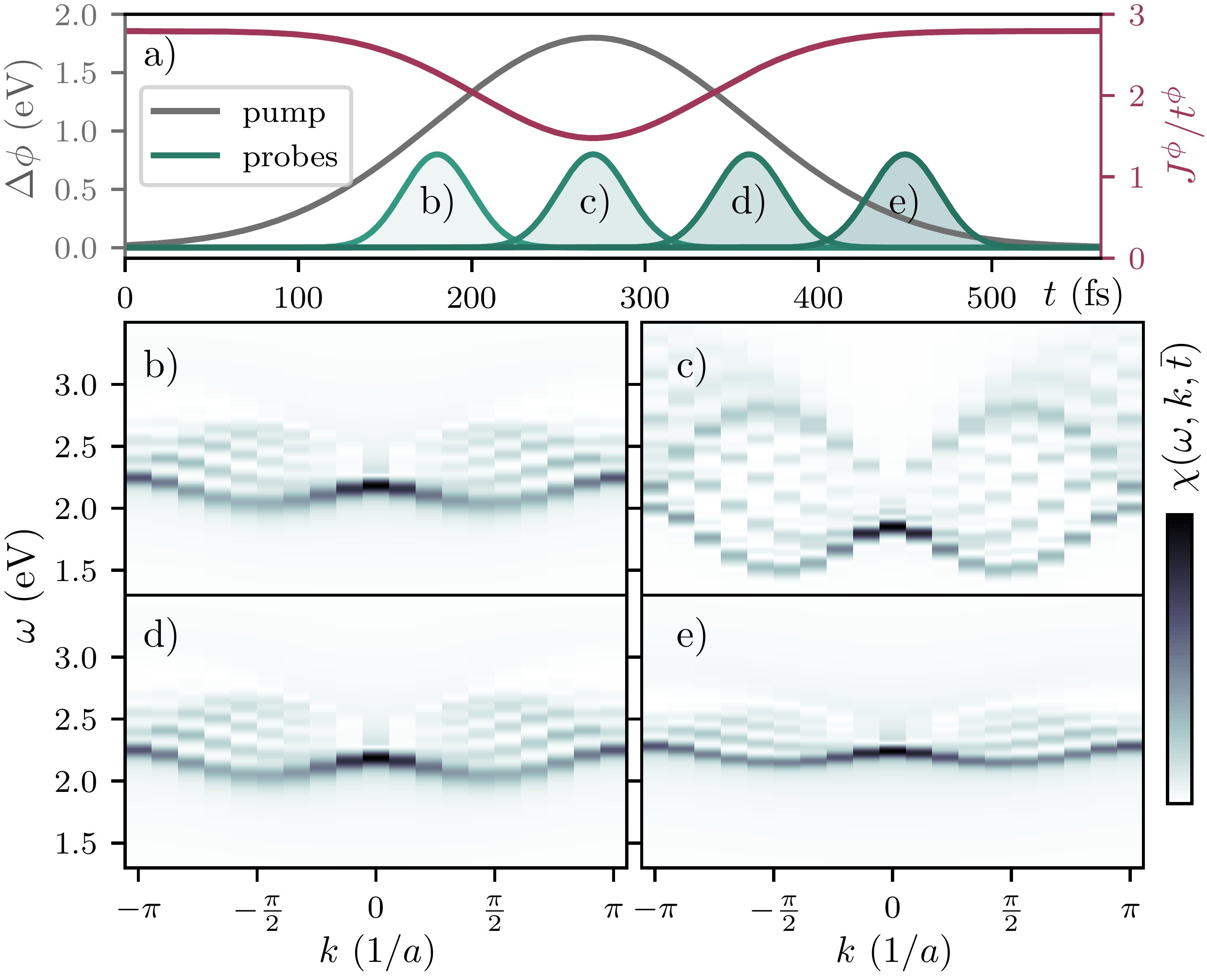}
    \vspace*{-6mm}
    \caption{a) Time-dependent potential field gradient $\Delta \phi(t)$ ($\Delta \phi^{max}=1.8$ eV, $\sigma_{pu}=90$ fs) and resulting change in $J/t$. b) - e) Time-dependent evolution of the spectral function $\chi(\omega, k, \bar{t})$ for different probe envelopes with $\sigma_{pr}=20$ fs and b) $\bar{t} = 180$ fs, c) $\bar{t} = 270$ fs, d) $\bar{t} = 360$ fs and e) $\bar{t} = 450$ fs for a chain length $L=18$.}
    \label{fig:td_Spectrum}
\end{figure}

\textit{Dynamic Control -- }
As elucidated in the previous sections, it is possible to achieve control over the parameters $t$ and $J$ of the effective $t-J$ model \equ{eq:H_t-J} both in the quasi-static and the Floquet regime. Here, we explicitly demonstrate the time-dependent modification of the spectrum for the quasi-static limit. We choose the time-dependent profile $\Delta \phi(t)$ to be Gaussian with variance $\sigma_{pu}=90$ fs and maximum value $\Delta \phi^{max}= 1.8$ eV, as depicted in \figu{fig:td_Spectrum}a). 

In \figu{fig:td_Spectrum}b)-e), we show the time evolution of $\chi(\omega, k, \bar t)$ during this time-dependent perturbation. At moderately strong values of the external field, the spectral function is not substantially changed with respect to equilibrium, even if quantitative differences can be observed, see \figu{fig:td_Spectrum}b). However, around the maximum $\Delta \phi^{max}$ one identifies a qualitative change in the spectrum, with the appearance of a well-defined upper orbiton branch  not present at lower values of the field, see \figu{fig:td_Spectrum}c). This is a signature of the transition from the fast to the slow spinon regime. Indeed, as shown in \figu{fig:td_Spectrum}a), the values of the ratio $J/t$ reached at $\Delta \phi^{max}$ are well below the supersymmetric point $J/t = 2$. As the pump strength decreases, the spectrum narrows into its equilibrium shape, which happens reversibly, see Figs.~\ref{fig:td_Spectrum}d) and e). This proves the possibility to dynamically control the qualitative shape of the spectral function and hence of the underlying collective excitations.

%%%%%%%%%%%%%%%%%%%%%%%%%%%%% CONCLUSIONS
%
\textit{Conclusions --} 
In conclusion, we showed that short pulses with realistic field amplitudes can be used to reversibly control the spin-orbital excitations in the quasi-one-dimensional compound Sr\textsubscript{2}CuO\textsubscript{3}. These pulses can substantially change the relative strength of the $t$ and $J$ parameters in the effective $t-J$ model which describes the spin and the orbital excitations in this material. This demonstrates control over the electronic fractionalization in solids between regimes where the spinon moves faster than the orbiton and the contrary one. More generally, the analysis illustrates that the spin-orbital physics in Mott insulators is a suitable platform to explore the reversible control of many-body dynamics in solids with strong laser fields. In particular, with the improved energy resolution available at new generation free electron lasers, spin-orbital excitations can be probed using trRIXS. Moreover, our analysis shows that a control of the low energy physics can be achieved by two different pathways, i.e., Floquet engineering and a sub-cycle control with strong THz transients, which can be seen as two limits of a more general control of exchange interactions with arbitrary time-dependent fields \cite{eckstein2017designing}. The spatially anisotropic character of the orbital-orbital interaction, as opposed to the isotropic spin-spin one, suggests a different physics in the two situations. Indeed, the light modulation of the orbital exchange interaction leads to a switch of the state of the system from a discrete minimum to another, each corresponding to distinguishable orbital orders \cite{Grandi2021_Orbital}. Besides the exemplary compound (Sr\textsubscript{2}CuO\textsubscript{3}) we considered, further Mott systems show spin-orbital separation, and we expect a nontrivial dynamic control in CaCu\textsubscript{2}O\textsubscript{3} \cite{PhysRevLett.114.096402} or Ca\textsubscript{2}CuO\textsubscript{3} \cite{PhysRevB.101.205117}, the light-control of multi-spinon excitations in Sr\textsubscript{2}CuO\textsubscript{3} \cite{Schlappa2018_NatComm} and the control of two-spinon and two-orbital excitations in the spin-Peierls compound TiOCl \cite{Glawion2011_PRL}.

%%%%%%%%%%%%%%%%%%%%%%%%%%%%% ACKNOWLEDGMENT
\textit{Acknowledgment - }
We thank Matteo Mitrano for useful discussions. This work was supported by the ERC Starting Grant No. 716648. The authors gratefully acknowledge the computational resources and support provided by the Regional Computing Center Erlangen (RRZE).

%%%%%%%%%%%%%%%%%%%%%%%%%%%%% BIBLIOGRAPHY
%\bibliography{biblio}{}

%merlin.mbs apsrev4-1.bst 2010-07-25 4.21a (PWD, AO, DPC) hacked
%Control: key (0)
%Control: author (0) dotless jnrlst
%Control: editor formatted (1) identically to author
%Control: production of article title (0) allowed
%Control: page (1) range
%Control: year (0) verbatim
%Control: production of eprint (0) enabled
%

%%%%%%%%%%%%%%%%%%%%%%%%%%%%% APPENDIX
\onecolumngrid
\numberwithin{equation}{subsection}
\begin{appendices}
  
\section*{Supplemental Material for: 'Ultrafast control of spin-orbital separation probed with time-resolved RIXS'}

The Supplemental Material is structured as follows: In \secu{App:A}, we discuss the derivation of the $t-J$ model (see Eq.~(1) in the main text) and its generalization to the driven case in the quasi-static and Floquet limits. In \secu{App:B}, we provide details about the derivation of the time-resolved RIXS intensity (Eq.~(4) in the main text) and compare it to ARPES.

\subsection{From the microscopic Hamiltonian to the effective $t-J$ model} \label{App:A}
The aim of this section is to derive a low energy Hamiltonian for the description of the orbital excitations in Sr$_2$CuO$_3$ (Eq.~(1) in the main text) and to generalize it to non-equilibrium conditions in the quasi-static and Floquet limits.

\subsubsection{Charge-transfer model} \label{charge_tr_mod}
The starting point is the ab initio charge-transfer Hamiltonian $\opr H_{c-t}=\opr H^1+\opr H^2+\opr H^3+\opr H^4$, which describes the interactions of the coppers and oxygen atoms in the quasi-one-dimensional compound Sr$_2$CuO$_3$ \cite{Wohlfeld2013}. The unit cell of this system contains one Cu site and three O atoms. One oxygen lies within the one-dimensional copper chain Cu-O-Cu-O-$\ldots$, and the other two lie above and below the Cu atom, respectively. $\opr H^1$ is the tight-binding contribution which includes a nearest-neighbor hopping between the copper and oxygen atoms, the chemical potential and the crystal field splitting between different orbitals
\begin{align} \label{eq:H1}
    \opr H^1= &-t_\sigma \sum_{j \sigma} \left( f_{j a \sigma}^\dagger f_{j x \sigma} - f_{j+1, a \sigma}^\dagger f_{j x \sigma} + H.c. \right) 
    - t_{\sigma o} \sum_{j, \sigma} \left( f_{j a \sigma}^\dagger f_{j yo^+ \sigma} - f_{j a \sigma}^\dagger f_{j yo^- \sigma} + H.c. \right)- t_{\pi} \sum_{j \sigma} \left( f_{j b \sigma}^\dagger f_{j z \sigma} - f_{j+1, b \sigma}^\dagger f_{j z \sigma} + H.c. \right) \nonumber \\
    &+ \varepsilon_a \sum_j n_{ja} + \varepsilon_b \sum_j n_{jb} + \Delta_x \sum_j \left( n_{jx}-n_{ja} \right) + \Delta_z \sum_j \left( n_{jz}-n_{jb} \right) \nonumber \\
    &+ \Delta_{yo} \sum_j \left( n_{jyo^+}-n_{ja} \right) + \Delta_{yo} \sum_j \left( n_{jyo^-}-n_{ja} \right) + \Delta_{xo} \sum_j \left( n_{jxo^+}-n_{jb} \right) + \Delta_{xo} \sum_j \left( n_{jxo^-}-n_{jb} \right) \, .
\end{align}
$\opr H^2$ is the onsite Coulomb repulsion on the copper sites, which includes intra- and interband Hubbard repulsion, Hund's interaction and pair-hopping terms
\begin{align} \label{eq:H2}
    \opr H^2=& \sum_{j,\sigma,\alpha<\beta} \left(U - 2J^{\alpha \beta}_H \right) n_{j \alpha \sigma} n_{j \beta \bar \sigma} + \sum_{j,\sigma,\alpha<\beta} \left(U - 3J^{\alpha \beta}_H \right) n_{j \alpha \sigma} n_{j \beta \sigma} + U \sum_{j,\alpha} n_{j \alpha \uparrow} n_{j \alpha \downarrow} \nonumber \\
    &- \sum_{j,\sigma,\alpha<\beta} J^{\alpha \beta}_H f_{j \alpha \sigma}^\dagger f_{j \alpha \bar \sigma} f_{j \beta \bar \sigma}^\dagger f_{j \beta \sigma} + \sum_{j,\sigma,\alpha<\beta} J^{\alpha \beta}_H f_{j \alpha \uparrow}^\dagger f_{j \alpha \downarrow}^\dagger f_{j \beta \downarrow} f_{j \beta \uparrow} \, .
\end{align}
Similarly, $\opr H^3$ describes the onsite Coulomb repulsion on the oxygen sites
\begin{align} \label{eq:H3}
    \opr H^3=& \sum_{j,\sigma,\mu<\nu} \left(U_p-2J^{\mu \nu}_H \right) n_{j \mu \sigma} n_{j \nu \bar \sigma} + \sum_{j,\sigma,\mu<\nu} \left(U_p-3J^{\mu \nu}_H \right) n_{j \mu \sigma} n_{j \nu \sigma} + U_p \sum_{j,\mu} n_{j \mu \uparrow} n_{j \mu \downarrow} \nonumber \\
    &- \sum_{j,\sigma,\mu<\nu} J^{\mu \nu}_H f_{j \mu \sigma}^\dagger f_{j \mu \bar \sigma} f_{j \nu \bar \sigma}^\dagger f_{j \nu \sigma} + \sum_{j,\sigma,\mu<\nu} J^{\mu \nu}_H f_{j \mu \uparrow}^\dagger f_{j \mu \downarrow}^\dagger f_{j \nu \downarrow} f_{j \nu \uparrow}\, .
\end{align}
Finally, $\opr H^4$ is the nearest-neighbor Coulomb repulsion
\begin{align} \label{eq:H4}
    \opr H^4=V_{dp} \sum_{j \mu \alpha} n_{j \alpha} \left( n_{j \mu} + n_{j \mu^+} + n_{j \mu^-} + n_{j+1, \mu} \right)\, . 
\end{align}
The two 3d Cu orbitals considered in the previous expressions are labeled as $a \equiv 3 \matn d_{x^2-y^2}$ and $b \equiv 3 \matn d_{zx}$, the 2p O orbitals within the chain are $x \equiv 2 \matn p_x$ and $z \equiv 2 \matn p_z$, and the 2p O orbitals above (below) the chain are $xo^+ \equiv 2 \matn p_x$ ($xo^- \equiv 2 \matn p_x$) and $yo^+ \equiv 2 \matn p_y$ ($yo^- \equiv 2 \matn p_y$). The indices $\alpha, \beta \in \{a,b\}$ describe the copper orbitals and $\mu,\nu \in \{x,z\}$ the oxygen orbitals. $f_{j \kappa \sigma}^\dagger$ creates a hole at site $j$ in orbital $\kappa \in \{a,b,x,z\}$ with spin $\sigma \in \{\uparrow,\downarrow\}$ ($\bar \sigma=-\sigma$), while the number operators are $n_{j \kappa}=n_{j \kappa \uparrow} + n_{j \kappa \downarrow}$ with $n_{j \kappa \sigma}=f_{j \kappa \sigma}^\dagger f_{j \kappa \sigma}$. In the hole representation and for the system in the ground state, there is a single hole in the unit cell which occupies the $a$ orbital. The other 3d Cu orbitals are not considered here as their characteristic dispersion relations are weaker. The relevant values of the parameters in Eqs. (\ref{eq:H1}) - (\ref{eq:H4}) are then $t_\sigma = 1.5$ eV, $t_\pi = 0.83$ eV, $t_{\sigma o} = 1.8$ eV, $t_{\pi o} = 1.0$ eV, $\Delta_x = 3.0$ eV, $\Delta_z = 2.2$ eV, $\Delta_{x o} = 2.5$ eV, $\Delta_{y o} = 3.5$ eV, $U=8.8$ eV, $U_p = 4.4$ eV, $J^b_H = J^{ab}_H = 1.2$ eV, $J^p_H = 0.83$ eV, $V_{dp} = 1.0$ eV, $\varepsilon_a = 0.0$ eV and $\varepsilon_b = -0.5$ eV, in accordance with \cite{Wohlfeld2013,Neudert2000}.

\subsubsection{$t-J$ model} \label{sec:t-Jmodel}
The charge-transfer Hamiltonian $\opr H_{c-t}$ provided in \secu{charge_tr_mod} belongs to the regime $t/U \ll 1$ and $t/\Delta < 1$, with $t$ any of the hoppings and $\Delta$ any of the crystal fields appearing in Eqs. (\ref{eq:H1}) - (\ref{eq:H4}), making the ground state a Mott insulator with a hole localized on the $a$ orbital on each copper site. Similarly, when a single orbital excitation is present, the hole is localized on the $b$ orbital. As outlined in \cite{Wohlfeld2013}, the second- and the fourth-orders of the perturbative expansion in the hoppings lead to the expression
\begin{align}\label{eq:strong_coupling}
    \bar{\opr{H}} = \opr H^{II} + \opr H^{IV}_a + \opr H^{IV}_b \, .
\end{align}
The second-order contribution has two important consequences: Firstly, it leads to a renormalization of the on-site energies of the orbitals; Secondly, it results in a renormalization of the hoppings within the chain, due to hybridization of the Cu orbitals with the oxygen atoms above and below the copper-oxygen chain. The first effect is given by
\begin{align} \label{eq:IIord}
    \opr H^{II} = \varepsilon_a \sum_j \tilde n_{ja} + \bar{\varepsilon}_b \sum_j \tilde n_{jb} \;,
\end{align}
where the tilde over the occupation operators implies that double occupations are prohibited in this strong coupling expansion and
\begin{align} \label{eq:cf_Undriven}
    \bar{\varepsilon}_b = \varepsilon_b + \frac{2 t_{\sigma o}^2}{\Delta_{y o}} + \frac{2 t_\sigma^2}{\Delta_x} - \frac{2 t_\pi^2}{\Delta_z} \;.
\end{align}
\equ{eq:IIord} is an effective crystal field between $a$ and $b$ orbitals. By introducing the pseudospin operators
\begin{align} \label{eq:pseodospin_operator_b}
    \sigma^z_j=\frac{1}{2}(\tilde n_{j b} - \tilde n_{j a}), \quad \sigma^+_j=\tilde f_{j b}^\dagger \tilde f_{j a}, \quad \sigma^-_j=\tilde f_{j a}^\dagger \tilde f_{j b} \;, 
\end{align}
one can rewrite \equ{eq:IIord} as
\begin{align}
    \opr H^{II} = \varepsilon_a \sum_j \left( \frac{1}{2} - \sigma^z_j \right) + \bar{\varepsilon}_b \sum_j \left( \frac{1}{2} + \sigma^z_j \right) \;.
\end{align}
The renormalization of the hopping affects only $\bar{t}_\sigma = \lambda_a t_\sigma$ with
\begin{align}
    \lambda_a=\frac{\Delta_{yo}-e_a}{\sqrt{2 t_{\sigma o}^2+(\Delta_{yo}-e_a)^2}} \, ,\qquad e_a=\frac{1}{2} \left( \Delta_{yo}-\sqrt{\Delta_{yo}^2+8 t_{\sigma o}^2} \right) \, .
\end{align}
For additional details about this derivation, we refer the reader to \cite{Wohlfeld2013}. 

The first fourth-order contribution to \equ{eq:strong_coupling} describes the superexchange process of two holes in the $a$ orbitals at two distinct copper atoms
\begin{align}\label{eq:IVord_a}
    \opr H^{IV}_a = J \sum_j \opr P_{j, j+1}  \left(\vec S_j \vec S_{j+1}-\frac{1}{4} \tilde n_j \tilde n_{j+1} \right) \,, 
\end{align}
where $\opr H^{IV}_a \equiv \opr H_s$ defined in the main text. Here, we introduced the spin operators
\begin{align} \label{eq:spin_operator}
    S^z_j=\frac{1}{2}(\tilde n_{j\uparrow} - \tilde n_{j\downarrow}), \quad S^+_j=\tilde f_{j\uparrow}^\dagger \tilde f_{j\downarrow}, \quad S^-_j=\tilde f_{j\downarrow}^\dagger \tilde f_{j\uparrow} \,,
\end{align}
and we defined the superexchange interaction $J = J^1 (1 + 2 R) > 0$, with
\begin{align} \label{eq:coefUndriven}
    J^1=\frac{4 \bar{t}_\sigma^4}{\left( \Delta_x+V_{dp} \right)^2 U} \, ,\qquad R=\frac{U}{2 \Delta_x + U_p} \, .
\end{align}
The operator $\opr P_{j, j+1} = ( 1/2  - \sigma_i^z) ( 1/2  - \sigma_{i+1}^z)$ projects out any orbital excitation along the bond $(j,j+1)$. The Hamiltonian \equ{eq:IVord_a} describes a Heisenberg antiferromagnetic interaction between nearest copper sites.

The second fourth-order contribution to \equ{eq:strong_coupling} describes the superexchange process in the presence of an orbital excitation, i.e. a single hole occupying the $b$ orbital. It reads
\begin{align} \label{eq:IVord_b}
    \opr H^{IV}_b = &\sum_{j} \left(\vec S_j \vec S_{j+1}+\frac{3}{4} \right) \left\{ \left[ \frac{r^{b1}}{2} \left(J^1+J^{b2} \right) + R^{b1} \sum_{\mu, \nu \in \{x,z\}} J^{b12}_{\mu \nu} \right] \left[\sigma_j^z \sigma_{j+1}^z -\frac{1}{4} \right]   
    + \left[ r^{b1} J^{b12}_{xz} +\frac{R^{b1}}{2} \sum_{\mu, \nu \in \{x,z\}} J^{b12}_{\mu \nu}
    \right] \left[ \sigma_j^+ \sigma_{j+1}^- +\sigma_j^- \sigma_{j+1}^+ \right] \right\} \nonumber \\ 
    +& \sum_{j} \left(\frac{1}{4} - \vec S_j \vec S_{j+1} \right) \left\{ \left[ \frac{r^{b2}}{2} \left(J^1+J^{b2} \right) + R^{b2} \sum_{\mu, \nu \in \{x,z\}} J^{b12}_{\mu \nu} \right] \left[\sigma_j^z \sigma_{j+1}^z -\frac{1}{4} \right] 
    - \left[ r^{b2} J^{b12}_{xz} +\frac{R^{b2}}{2} \sum_{\mu, \nu \in \{x,z\}} J^{b12}_{\mu \nu}
    \right] \left[ \sigma_j^+ \sigma_{j+1}^- +\sigma_j^- \sigma_{j+1}^+ \right] \right\} 
\end{align}
with
\begin{alignat}{3} \label{eq:coefCal_Undriv}
    &J^{b2}=\frac{4 t_\pi^4}{\left( \Delta_z+V_{dp} \right)^2 U}\, ,
    &&r^{b1}=\frac{U}{U-3J_H^b}\, ,\quad 
    &&R^{b1}=\frac{U}{\Delta_x+ \Delta_z + U_p-3J^p_H}\, ,
    \\
    &J_{\mu \nu}^{b12}=\frac{2 t_\pi^2 \bar{t}_\sigma^2}{\left( \Delta_\mu+V_{dp} \right) U \left( \Delta_\nu+V_{dp} \right)}\, , \quad
    &&r^{b2}=\frac{U}{U-J_H^b }\, ,
    &&R^{b2}=\frac{U}{\Delta_x+ \Delta_z + U_p-J^p_H}\,.
\end{alignat}
\equ{eq:IVord_b} is a spin-orbital Hamiltonian of the Kugel-Khomskii type. 

The Hamiltonian \equ{eq:strong_coupling} describes the low-energy spin-orbital dynamics of the quasi-one-dimensional compound Sr$_2$CuO$_3$. Given the dimensionality of the system, spin and pseudospin operators Eqs. (\ref{eq:spin_operator}) and (\ref{eq:pseodospin_operator_b}) can be conveniently rewritten through the Jordan-Wigner transformation
\begin{align} 
    S_j^+ =\exp \left( -i \pi \sum_{l=1}^{j-1} n_{l \alpha} \right) \alpha_j^\dagger\,,  \quad
    S_j^- = \alpha_j \exp \left(i \pi \sum_{l=1}^{j-1} n_{l \alpha} \right),  \quad
    S_j^z=n_{j \alpha}-\frac{1}{2} \, , \label{eq:jord_wign_spin} \\ 
    \sigma_j^+ =\exp \left( -i \pi \sum_{l=1}^{j-1} n_{l \beta} \right) \beta_j^\dagger\,, \quad
    \sigma_j^-= \beta_j \exp \left(i \pi \sum_{l=1}^{j-1} n_{l \beta} \right), \quad 
    \sigma_j^z=n_{j \beta}-\frac{1}{2} \, ,
\end{align}
where $\alpha_j$ ($\beta_j$) is a fermionic operator annihilating a spinon (pseudospinon) on site $j$ and $n_{j \alpha}$ ($n_{j \beta}$) is the spinon (pseudospinon) number operator. As discussed in \cite{Wohlfeld2013}, a spinon and a pseudospinon cannot be on the same site, leading to the constraint $n_{j \alpha} + n_{j \beta} \le 1$ for each site $j$, and it is assumed that the hopping of the pseudospinon does not change the spin of the hole in the $b$ orbital.

Finally, by defining the fermionic operators
\begin{align}
    \tilde p_{j \uparrow}= \beta_j^\dagger \, , \qquad \tilde p_{j \uparrow}= \beta_j^\dagger \alpha_j \exp \left( i \pi \sum_{l=1}^{j-1} n_{l \alpha} \right),
\end{align}
which act in a Hilbert space without double occupations and introducing back the spin operators \equ{eq:jord_wign_spin}, the Hamiltonian \equ{eq:strong_coupling} takes the form
\be \label{eq:H_t-J_app}
    \bar{\opr{H}} \equiv \opr H_{t-J} =  -t \sum_{j, \sigma} \left( p_{j \sigma}^\dagger p_{j+1 \sigma} + h.c. \right)  - E \sum_j  \tilde n_j +  J \sum_j  \left( \vec S_j \vec S_{j+1} - \frac{1}{4} \tilde n_j \tilde n_{j+1}\right),
\ee
with
\begin{align} \label{eq:eff_mic_param_Undriv}
    J &= J^1 (1 + 2 R) \, , \\
    t &= \frac{1}{4} \left[ 2 (3 r^{b1}+r^{b2}) J^{b12}_{xz} + (3 R^{b1}+R^{b2}) \sum_{\mu, \nu \in \{x,z\}} J^{b12}_{\mu \nu} \right] , \\
    E &= \bar{\varepsilon}_b - \left[ \frac{r^{b1}}{2} \left(J^1+J^{b2} \right) + R^{b1} \sum_{\mu, \nu \in \{x,z\}} J^{b12}_{\mu \nu} \right] \;.
\end{align}
\equ{eq:H_t-J_app} is equivalent to Eq.~(1) of the main text.

\subsubsection{Quasi-static limit}
In this section, we analyze how the parameters of the effective $t-J$ model \equ{eq:eff_mic_param_Undriv} change under the action of an electric field included as a scalar potential $\phi_{j \kappa}$, which we assume slowly varying, i.e. $\Omega = 0 $. The charge-transfer Hamiltonian, in this case, has an additional contribution
\be \label{eq:H_micro}
    \opr H_{c-t}^\phi = \opr H_{c-t} + \sum_{j,\kappa} \phi_{j \kappa} n_{j \kappa} \, .
\ee
Considering the same derivation sketched in \secu{sec:t-Jmodel}, one arrives at an expression for the $t-J$ model akin to \equ{eq:H_t-J_app}, with new parameters
\begin{align} \label{eq:eff_mic_param_qstat}
    J^\phi&=\sum_{s \in \{+,-\}} J_s^1 r_s^a+R^a \left(\sqrt{J_+^1}+\sqrt{J_-^1}\right)^2 \, , \\
    t^\phi &= \sum_{s \in \{+,-\}} \left[ \frac{3 r^{b1}_s+r^{b2}_s}{8} \left( J^{b12}_{xz,s}+J^{b12}_{zx,s} \right) + \frac{3 R^{b1}+R^{b2}}{8} \sum_{\mu, \nu \in \{x,z\}} J^{b12}_{\mu \nu, s} \right] , \\
    E^\phi &= \bar{\varepsilon}_b^\phi - \sum_{s \in \{+,-\}} \left[ \frac{r^{b1}_s}{2} \left(J^1_s+J^{b2}_s \right) + R^{b1} \sum_{\mu, \nu \in \{x,z\}} J^{b12}_{\mu \nu, s} \right] 
\end{align}
with
\begin{alignat}{3} \label{eq:coefCal_quasiStat}
    &J_{\pm}^1=\frac{2 \bar{t}_\sigma^4}{\left( \Delta_x+V_{dp} \pm \Delta \phi \right)^2 U} \,,
    && r_{\pm}^a=\frac{U}{U \pm 2 \Delta \phi}\,,
    && R^a=\frac{U}{2 \Delta_x + U_p} \, ,\\
    &
    J_{\pm}^{b2}=\frac{2 t_\pi^4}{\left( \Delta_z+V_{dp} \pm \Delta \phi \right)^2 U}\, ,
    &&r_{\pm}^{b1}=\frac{U}{U-3J_H^b \pm 2 \Delta \phi}\, ,\quad 
    &&R^{b1}=\frac{U}{\Delta_x+ \Delta_z + U_p-3J^p_H}\, ,
    \\
    &J_{\mu \nu, \pm}^{b12}=\frac{t_\pi^2 \bar{t}_\sigma^2}{\left( \Delta_\mu+V_{dp} \pm \Delta \phi\right) U \left( \Delta_\nu+V_{dp} \pm \Delta \phi \right)}\, , \quad
    &&r_{\pm}^{b2}=\frac{U}{U-J_H^b \pm 2 \Delta \phi}\, ,
    &&R^{b2}=\frac{U}{\Delta_x+ \Delta_z + U_p-J^p_H}\, ,
\end{alignat}
and
\begin{align} \label{eq:cf_phi}
    \bar{\varepsilon}_b^\phi = \varepsilon_b + \frac{2 t_{\sigma o}^2}{\Delta_{y o}} + t_\sigma^2 \left( \frac{1}{\Delta_x+\Delta \phi} + \frac{1}{\Delta_x-\Delta \phi} \right) - t_\pi^2 \left( \frac{1}{\Delta_z+\Delta \phi} + \frac{1}{\Delta_z-\Delta \phi} \right) \;.
\end{align}
The time-dependency of the problem is provided by the substitution $\Delta \phi \rightarrow \Delta \phi (t)= \phi_{j \mu}(t)-\phi_{j \alpha}(t) \equiv \phi_{j+1 \alpha}(t)-\phi_{j \mu}(t)$ for $\alpha \in \{a,b\}$ and $\mu \in \{x,z\}$.

In the main text, we did not take into account the renormalization of the crystal field \equ{eq:cf_phi} provided by the scalar potential, considering instead its expression in equilibrium \equ{eq:cf_Undriven}, as it only shifts the energies.

\subsubsection{Floquet limit}
The effect of a driving laser on the charge-transfer Hamiltonian is included via the Peierls substitution, which modifies the hopping between two nearest neighbor sites $i$ and $j$ as $(t_\sigma)_{ij} \rightarrow t_\sigma \exp (i e a A_{ij} (t)/2 )$ and $(t_\pi)_{ij} \rightarrow t_\pi \exp (i e a A_{ij} (t)/2 )$. Here $A_{ij}(t)$ is a time-dependent vector potential which we assume to be directed along the chain direction, so it does not influence the hoppings $t_{\sigma o}$ and $t_{\pi o}$. Considering an electric field $E(t) = E_0 \sin (\Omega t)$, the corresponding vector potential reads $A_{ij} (t) = - E_0 (i - j) \cos (\Omega t) / \Omega$. (Note that the coupling of the vector potential via the Peierls phase is gauge equivalent to coupling the local density to the scalar potential [Eq.~\eqref{eq:H_micro}], which is used to describe the quasi-static limit, see, e.g., Ref.~\cite{Boykin2001}.) This way, the charge-transfer Hamiltonian $\opr H_{c-t}^{\text{Fl}} (t)$ becomes periodic in time with period $T = 2 \pi / \Omega$ ($\opr H_{c-t}^{\text{Fl}} (t) = \opr H_{c-t}^{\text{Fl}} (t + T)$). The solution of the Schr\"odinger equation $\vert \psi (t) \rangle$ can be rewritten, by using the Floquet theorem
\begin{align}
    %AM: as discussed with FG it should be - \epsilon_\alpha
    \Ket{ \psi (t)} = e^{-i \epsilon_\alpha t } \Ket{ \psi_\alpha (t)} \, ,
    %\Ket{ \psi (t)} = e^{i \epsilon_\alpha t } \Ket{ \psi_\alpha (t)} \, ,
\end{align}
with $\vert \psi_\alpha (t) \rangle = \vert \psi_\alpha (t + T) \rangle$ time periodic and $\epsilon_\alpha$ a Floquet quasi-energy defined up to integer multiples of $\Omega$. $\vert \psi_\alpha (t) \rangle$ can be expanded in Fourier series as $\vert \psi_\alpha (t) \rangle = \sum_{n=-\infty}^\infty \exp (-i \Omega n t) \, \vert \psi_{\alpha,n} \rangle$, where $\vert \psi_{\alpha,n} \rangle$ is the component of the wave function in the n-th Floquet sector. In this basis, the Schr\"odinger equation reads
\begin{align}
    (\epsilon_\alpha + n \Omega) \vert \psi_{\alpha,n} \rangle = \sum_{n'} (\opr H_{c-t}^{\text{Fl}})_{n-n'} \vert \psi_{\alpha,n'} \rangle 
\end{align}
with $(\opr H_{c-t}^{\text{Fl}})_n$ the n-th Fourier component of the Hamiltonian $(\opr H_{c-t}^{\text{Fl}})_n = 1/T \int_0^T \ dt \exp(i \Omega n t) \, \opr H_{c-t}^{\text{Fl}}(t)$. Thereby, only the tight-binding contribution $\opr H^1$ of the charge-transfer Hamiltonian is changed 
\begin{align} \label{eq:H1floquet}
    (\opr H^1)_n = &-t_\sigma J_n (\mathcal{E} ) \sum_{j \sigma} \left( f_{j a \sigma}^\dagger f_{j x \sigma} - (-1)^n f_{j+1, a \sigma}^\dagger f_{j x \sigma} + (-1)^n f_{j x \sigma}^\dagger f_{j a \sigma} - f_{j x \sigma}^\dagger f_{j+1, a \sigma} \right) \nonumber \\
    &- t_{\pi} J_n (\mathcal{E} ) \sum_{j \sigma} \left( f_{j b \sigma}^\dagger f_{j z \sigma} - (-1)^n f_{j+1, b \sigma}^\dagger f_{j z \sigma} + (-1)^n f_{j z \sigma}^\dagger f_{j b \sigma} - f_{j z \sigma}^\dagger f_{j+1, b \sigma} \right) \nonumber \\
    &- t_{\sigma o} \sum_{j, \sigma} \left( f_{j a \sigma}^\dagger f_{j yo^+ \sigma} - f_{j a \sigma}^\dagger f_{j yo^- \sigma} + H.c. \right) + \varepsilon_a \sum_j n_{ja} + \varepsilon_b \sum_j n_{jb} \nonumber \\
    & + \Delta_x \sum_j \left( n_{jx}-n_{ja} \right) + \Delta_z \sum_j \left( n_{jz}-n_{jb} \right) + \Delta_{yo} \sum_j \left( n_{jyo^+}-n_{ja} \right) + \Delta_{yo} \sum_j \left( n_{jyo^-}-n_{ja} \right) \, ,
\end{align}
where $J_n (\mathcal{E})$ is the n-th Bessel function of the first kind and $\mathcal{E} = e \tilde a E_0/ (2 \Omega)$ is the Floquet parameter. Thus, the hopping allows the system to change Floquet sectors by the emission or the absorption of photons. Starting from the effective time-independent Floquet Hamiltonian, one can follow a similar procedure as presented in the previous sections, leading to a $t-J$ model with parameters
\begin{align} \label{eq:eff_mic_param_floquet}
    J^{\opr E,\Omega} &= \sum_{n,l,k} J^1_{nlk} (1+ R^a_{nlk}),  \\
    t^{\opr E,\Omega} &= \sum_{n,l,k} \left[ \frac{3 r^{b1}_l + r_l^{b2}}{8} \left(J^{b12}_{xz,nlk} + J^{b12}_{zx,nlk} \right) + \frac{3 R^{b1}_{nlk} + R^{b2}_{nlk}}{8} \left( (-1)^l J^{b12}_{xx, nlk} + (-1)^l J^{b12}_{zz, nlk} + J^{b12}_{xz, nlk}+ J^{b12}_{zx, nlk}\right) \right] \, , \\
    E^{\opr E,\Omega} &= \bar{\varepsilon}_b^{\opr E,\Omega} -\sum_{n,l,k} \left[ \frac{r^{b1}_l}{2} \left(J^1_{nlk} + J^{b2}_{nlk} \right) + R^{b1}_{nlk} \left( J^{b12}_{xx, nlk} + J^{b12}_{zz, nlk} + (-1)^l J^{b12}_{xz, nlk}+(-1)^l J^{b12}_{zx, nlk}\right) \right] \;,
\end{align}
where
\begin{alignat}{3} \label{eq:coefCal_floquet}
    &J^1_{nlk} = \frac{4 \bar t_\sigma^4 J_n (\opr E) J_{l-n} (\opr E) J_{l-k} ( \opr E) J_{k} ( \opr E)}{( \Delta_x +V_{dp} + n \omega ) ( U + l \omega ) ( \Delta_x + V_{dp} + k \omega )} \, , \quad
    && \hphantom{r^{b1}_l = \frac{U + l \omega}{ U - 3 J_H + l \omega}\, ,} 
    \quad && R^a_{nlk} = (-1)^{n+k} \frac{U + l \omega}{2 \Delta_x + U_p + l \omega} \left(1+(-1)^l\right) \, , \\
    & J^{b2}_{nlk} = \frac{4 t_\pi^4 J_n (\opr E) J_{l-n} (\opr E) J_{l-k} ( \opr E) J_{k} ( \opr E)}{( \Delta_z +V_{dp} + n \omega ) ( U + l \omega ) ( \Delta_z + V_{dp} + k \omega )} \, ,
    &&r^{b1}_l = \frac{U + l \omega}{ U - 3 J_H + l \omega}\, , 
    &&R^{b1}_{nlk} = (-1)^{n+k} \frac{U + l \omega}{ \Delta_x + \Delta_z + U_p - 3 J_H^p + l \omega} \, ,
    \\
    &J^{b12}_{\mu \nu,nlk} = \frac{2 t_\pi^2 \bar t_\sigma^2 J_n (\opr E) J_{l-n} (\opr E) J_{l-k} ( \opr E) J_{k} ( \opr E)}{( \Delta_\mu +V_{dp} + n \omega ) ( U + l \omega ) ( \Delta_\nu + V_{dp} + k \omega )} \;,
    &&r^{b2}_l = \frac{U + l \omega}{ U - J_H + l \omega}\, ,
    &&R^{b2}_{nlk} = (-1)^{n+k} \frac{U + l \omega}{ \Delta_x + \Delta_z + U_p - J_H^p + l \omega}
\end{alignat}
and
\begin{align}
    \bar{\varepsilon}_b^{\opr E,\Omega} = \varepsilon_b + \frac{2 t_{\sigma o}^2}{\Delta_{y o}} + t_\sigma^2 \sum_{n=-\infty}^{+\infty} \frac{J_{n}^2 (\opr E)}{\Delta_x + n \Omega} - t_\pi^2 \sum_{n=-\infty}^{+\infty} \frac{J_{n}^2 (\opr E)}{\Delta_z + n \Omega} \;.
\end{align}

%%%%%%%%%%%%%%%%%%% trRIXS
%
\subsection{Theory of time-resolved RIXS} \label{App:B}
trRIXS is a photon-in photon-out technique which can be described as a second order X-ray scattering process. Before the arrival of the pump, we assume the system to be in the ground state $\Ket{\Psi_0}$ of $\opr H_{c-t}$, i.e. a state in which the holes occupy the $a$ orbitals of the coppers. The incoming photon with energy $\omega_i$, momentum $k_i$ and polarization $\vec e_i$ can excite the $a$ hole to the $2$p core state of a copper atom. Within the dipole approximation, this process can be described by the dipole transition operator $\opr D_{j \vec e_i}(t)$. The highly unstable intermediate state with the core-hole decays within few femtoseconds in one of the $3 \matn d$ orbital by emitting a photon with energy $\omega_f$, momentum $k_f$ and polarization $\vec{e}_f$ (described by $\opr D^\dagger_{j \vec e_f}(t)$). The overall process is called direct RIXS, which is the dominant scattering channel since the transition $2 \matn p \leftrightarrow 3 \matn d$ is dipole allowed \cite{Marra2016}. 

If we turn on the pump, the system is no longer in equilibrium and we describe it with 
\begin{align}
    \bar{\opr H}_0 (t) = \opr H_0 (t) + \sum_{\substack{k_i, \vec{e}_i \\ k_f, \vec{e}_f}} \Big( \omega_i a^\dagger_{k_i, \vec{e}_i} a_{k_i, \vec{e}_i} + \omega_f a^\dagger_{k_f, \vec{e}_f} a_{k_f, \vec{e}_f} \Big) \, ,
\end{align}
where $\opr H_0 (t)$ is the time-dependent $\opr H_{c-t}$ and the last contribution is the free photon Hamiltonian $\opr H_{ph}$.
 
In the interaction (Dirac) picture the states evolve as $\bar{\opr U}_0(t,t_0)| \Psi_0 \rangle \otimes | \Psi_{ph} \rangle$, where $\bar{\opr U}_0(t,t_0)=\opr T_D \exp \left(-i \int_{t_0}^t \matn d t' \bar{\opr{H}}_0(t')\right)$ is the time evolution operator with Dyson’s time-ordering operator $\opr T_D$. For the later use we analogously define time evolution operators for the photonic and electronic subsystem $\opr U_{ph} (t,t_0)= \exp \left(-i \opr H_{ph} (t-t_0)\right)$  and $\opr U_0(t,t_0)=\opr T_D \exp \left(-i \int_{t_0}^t \matn d t' \opr H_0(t')\right)$. The photonic states in the weak-probe limit can be approximated as $| \Psi_{ph} \rangle \approx a^\dagger_{{k}_i \vec{e}_i} \ket{0}$. The probe is generally much weaker than the pump, thus we can treat its contribution as a perturbation $\opr H'(t)$ of the full Hamiltonian $\opr H(t)= \bar{\opr H}_0(t)+ \opr H'(t)$. By expanding to second order the time evolution operator reads
\begin{align}
    \opr U(t,t_0) & = \opr T_D \exp \left(-i \int_{t_0}^t \matn d t' \opr H(t')\right) \\
    & \approx \bar{\opr U}_0(t,t_0)-i \int_{t_0}^t \matn d t_1 \bar{\opr U}_0(t,t_1) \opr H'(t_1) \bar{\opr U}_0(t_1,t_0) - \int_{t_0}^t \matn d t_2 \int_{t_0}^{t_2} \matn d t_1 \bar{\opr U}_0(t,t_2) \opr H'(t_2) \bar{\opr U}_0(t_2,t_1) \opr H'(t_1) \bar{\opr U}_0(t_1,t_0) \, . 
\end{align}
In the experiment one can measure the flux of photons with given momentum, polarization and energy
\begin{align}
    I_{ k_f \vec e_f} & = \lim_{t \rightarrow \infty} \Braket{\bar{\opr U}_0^\dagger(t,t_0) a_{ k_f \vec e_f}^\dagger a_{ k_f \vec e_f} \bar{\opr U}_0(t,t_0)} \nonumber \\
    & \approx I_{ k_f \vec e_f}^{(0)}+I_{ k_f \vec e_f}^{(2)} + I_{ k_f \vec e_f}^{(4)} \; ,\label{eq:J_orderByOrder}
\end{align}
which can be calculated order by order. Note that in \equ{eq:J_orderByOrder}, just the even terms survive since they are the only ones that guarantee the same amount of photon creation and annihilation operators in the expectation value. The zeroth order corresponds to elastic reflection of the photon and the second order to time-resolved X-ray absorption spectroscopy (trXAS). The fourth order contains the actual RIXS spectrum, where we have to include an incoming and outgoing process on each branch of the Keldysh contour. Therefore, we decompose the probe Hamiltonian into $\opr H'(t)=\opr H_{in}(t)+\opr H_{out}(t) + H.c.$ and define them as 
\begin{align}
    \opr H_{in}(t)=s(t) \frac{1}{\sqrt{L}} \sum_{j {q} \vec e} \opr D_{j \vec e}(t) e^{i q j}  a_{ q \vec e} ,\qquad \opr H_{out}(t)= \frac{1}{\sqrt{L}} \sum_{j {q} \vec e} a^\dagger_{ q \vec e}  e^{-i q j} \opr D^\dagger_{j \vec e}(t) \; ,
\end{align}
where $s(t) \equiv s(t,\bar t)$ is the probe envelope function defined in the main text. In direct RIXS the core electron is excited directly into the valence band. Thus, the local dipole transition operator is $\opr D_{j \vec e}(t)=\sum_{\sigma \alpha \rho} M^{\vec e}_{\alpha \rho}(t) f^\dagger_{j \alpha \sigma} f_{j \rho \sigma}$, where $f_{j \alpha \sigma}$ ($f_{j \rho \sigma}$) annihilates an electron at site $j$ with spin $\sigma$ in valence  $\alpha$ (core $\rho$) orbital. $M^{\vec e}_{\alpha \rho}(t)$ denotes the associated matrix element with photon polarization $\vec e$, which, in non-equilibrium, can be time-dependent \cite{Chen2019}. If we only collect the relevant terms for trRIXS we get
\begin{align}
     I_{RIXS} \coloneqq I_{ k_f \vec e_f}^{(4)} &= \int_{t_0}^\infty \matn d t'_2 \int_{t_0}^{t'_2} \matn d t'_1 \int_{t_0}^\infty \matn d t_2 \int_{t_0}^{t_2} \matn d t_1 \Big \langle \bar{\opr U}_0(t_0,t'_1) \opr H^\dagger_{in}(t'_1) \bar{\opr U}_0(t'_1,t'_2) \opr H^\dagger_{out}(t'_2) \bar{\opr U}_0(t'_2,\infty) a_{ k_f \vec e_f}^\dagger \nonumber \\
    &\hphantom{==} \times a_{ k_f \vec e_f}\bar{\opr U}_0(\infty,t_2) \opr H_{out}(t_2) \bar{\opr U}_0(t_2,t_1) \opr H_{in}(t_1) \bar{\opr U}_0(t_1,t_0) \Big \rangle \\
    & = \frac{1}{L^2} \sum_{\substack{m j \\ m' j'}} \sum_{\substack{ q'_2  q'_1  q_2  q_1 \\ \vec e'_2 \vec e'_1 \vec e_2 \vec e_1}} \int_{t_0}^\infty \matn d t'_2 \int_{t_0}^{t'_2} \matn d t'_1 \int_{t_0}^\infty \matn d t_2 \int_{t_0}^{t_2} \matn d t_1  s(t'_1) s(t_1) \nonumber \\
    &\hphantom{==}\times \Big \langle \bar{\opr U}_0(t_0,t'_1) a^\dagger_{ q'_1  e'_1}  e^{-i  q'_1 m'} \opr D_{{m'} \vec e'_1}^\dagger(t'_1) \bar{\opr U}_0(t'_1,t'_2) \opr D_{j' \vec e'_2}(t'_2) e^{i  q'_2 j'}  a_{ q'_2 \vec e'_2} \bar{\opr U}_0(t'_2,\infty) a_{ k_f \vec e_f}^\dagger \nonumber \\
    &\hphantom{==}\times a_{ k_f \vec e_f} \bar{\opr U}_0(\infty,t_2) a^\dagger_{ q_2 \vec e_2}  e^{-i  q_2 m} \opr D_{{m} \vec e_2}^\dagger(t_2) \bar{\opr U}_0(t_2,t_1) \opr D_{j \vec e_1}(t_1) e^{i  q_1 j}  a_{ q_1 \vec e_1} \bar{\opr U}_0(t_1,t_0) \Big \rangle\,.\label{RIXS_ph_e_cont}
\end{align}
By separating the average value into the electronic and photonic part, we realize that for the photonic part it has to hold
\begin{align} \label{RIXS_ph_cont}
    \Bra{0} a_{ k_i \vec e_i} \opr U_{ph}(t_0,t'_1) a^\dagger_{ q'_1 \vec e'_1} \opr U_{ph}(t'_1,t'_2)  a_{ q'_2 \vec e'_2} \opr U_{ph}(t'_2,\infty) a_{ k_f \vec e_f}^\dagger a_{ k_f \vec e_f} \opr U_{ph}(\infty,t_2) a^\dagger_{ q_2 \vec e_2} \opr U_{ph}(t_2,t_1) a_{ q_1 \vec e_1} \opr U_{ph}(t_1,t_0) a_{ k_i \vec e_i}^\dagger \Ket{0} = \nonumber \\
    =\delta_{ k_i  q'_1} \delta_{\vec e_i \vec e'_1} \delta_{ k_i  q_1} \delta_{\vec e_i \vec e_1} \delta_{ k_f  q'_2} \delta_{\vec e_f \vec e'_2} \delta_{ k_f  q_2} \delta_{\vec e_f \vec e_2} e^{i \omega_i(t'_1-t_1)-i \omega_f(t'_2-t_2)} \, .
\end{align}
We assume that in the intermediate state the core hole does not move, i.e. $m=j$ and $m'=j'$. Substituting that, \equ{RIXS_ph_cont} and $ k =  k_f - k_i$ into \equ{RIXS_ph_e_cont} yields
\begin{align}
    I_{RIXS}=& \frac{1}{L} \sum_{j j'} \int_{t_0}^\infty \matn d t'_2 \int_{t_0}^{t'_2} \matn d t'_1 \int_{t_0}^\infty \matn d t_2 \int_{t_0}^{t_2} \matn d t_1  s(t'_1) s(t_1) e^{-i  k (j - j')} e^{i \omega_i(t'_1-t_1)-i \omega_f(t'_2-t_2)} \nonumber \\
    &\times \Big \langle \opr U_0(t_0,t'_1) \opr D_{j' \vec e_i}^\dagger(t'_1) \opr U_0(t'_1,t'_2) \opr D_{j' \vec e_f}(t'_2) \opr U_0(t'_2,t_2) \opr D_{j \vec e_f}^\dagger(t_2) \opr U_0(t_2,t_1) \opr D_{j \vec e_i}(t_1) \opr U_0(t_1,t_0) \Big \rangle \, .\label{eq:four-point_cor}
\end{align}
By introducing the local RIXS operator
\begin{align}
    \opr B_{j \vec e_i \vec{e_f}} \coloneqq \opr D_{j \vec e_f}^\dagger(t_1)  \frac{1}{(\omega_f -\opr H_0+i \Gamma)} \opr D_{j \vec e_i}(t_1)
\end{align}
with $1/\Gamma$ the core-hole lifetime \cite{PhysRevLett.109.117401}, we can implement the ultra-short core-hole lifetime approximation \cite{Ament_PRB_2007}, which reduces \equ{eq:four-point_cor} to a two point correlation function
\begin{align} \label{eq:Irixs_loc_O}
    I_{RIXS} \approx \frac{1}{L} \sum_{j'=0}^{L-1} \int_{t_0}^\infty \matn d t'_1 \int_{t_0}^{\infty} \matn d t_1  s(t'_1) s(t_1) e^{-i \omega( t_1-t'_1)} e^{i k j'} \Bra{\Psi_0} \opr U_0(t_0,t'_1) \opr B_{j' \vec e_i \vec e_f}^\dagger \opr U_0(t'_1,t_1) \opr B_{0 \vec e_i \vec e_f} \opr U_0(t_1,t_0) \Ket{\Psi_0} \;,
\end{align}
where $\omega \coloneqq \omega_i-\omega_f$. The time evolution operator $\opr U_0(t',t)$ now acts in the low energy sector, i.e. there are no charge excitations. It is, thus, determined by the $t-J$ or Heisenberg Hamiltonian in the quasi-static or Floquet limit. Since we work with periodic boundary conditions (PBC), in \equ{eq:Irixs_loc_O} we assume, without any loss of generality, the action of the first $\opr B_{j \vec e_i \vec{e_f}}$ on site $j=0$. The local RIXS operators can than be expanded in the basis of the local hole spin and orbital as
\begin{align}
    \opr B_{j \vec e_i \vec e_f}=\sum_{\sigma \sigma'}  B^{\sigma \sigma'}_{\vec e_i \vec e_f} f_{j b\sigma}^\dagger f_{j a\sigma'} \; .
\end{align}
The RIXS matrix elements $B^{\sigma \sigma'}_{\vec e_i \vec e_f}$ can be calculated following \cite{Marra2016}.

By considering that the derived $t-J$ Hamiltonian conserves the spin of the hole, we can write \equ{eq:Irixs_loc_O} as
\begin{align} \label{eq:Irixs_loc_O2}
    I_{RIXS}=& \frac{1}{L} \int_{t_0}^\infty \matn d t'_1 \int_{t_0}^{\infty} \matn d t_1  s(t'_1) s(t_1) e^{-i \omega ( t_1-t'_1)} \Big[ \Bra{\Psi_0} \opr U_0(t_0,t'_1) \sum_{j'} e^{i k j'} \sum_{\sigma'} B^{\uparrow \sigma' *}_{\vec e_i \vec e_f} f_{j' a \sigma'}^\dagger f_{j' b \uparrow} \opr U_0(t'_1,t_1) \sum_{\sigma} B^{\uparrow \sigma}_{\vec e_i \vec e_f} f_{0 b \uparrow}^\dagger f_{0 a \sigma'} \opr U_0(t_1,t_0) \Ket{\Psi_0} \nonumber \\
    &+ \Bra{\Psi_0} \opr U_0(t_0,t'_1) \sum_{j'} e^{i k j'} \sum_{\sigma'} B^{\downarrow \sigma *}_{\vec e_i \vec e_f} f_{j' a \sigma}^\dagger f_{j' b \downarrow} \opr U_0(t'_1,t_1) \sum_{\sigma} B^{\downarrow \sigma}_{\vec e_i \vec e_f} f_{0 b \downarrow}^\dagger f_{0 a \sigma} \opr U_0(t_1,t_0) \Ket{\Psi_0} \Big].
\end{align}
If we now apply the transformations explained in App.\ref{App:A} to obtain the effective $t-J$ Hamiltonian, we get for one of the local RIXS operators
\begin{align} \label{eq:B_f_to_p}
    \sum_\sigma B^{\uparrow \sigma}_{\vec e_i \vec e_f} f_{0 b \uparrow}^\dagger f_{0 a \sigma} =B^{\uparrow \uparrow}_{\vec e_i \vec e_f} p_{0 \uparrow} + B^{\uparrow \downarrow}_{\vec e_i \vec e_f} p_{0 \downarrow}\, ,
\end{align}
whereby the RIXS matrix elements generally obey $B^{\uparrow \uparrow}_{\vec e_i \vec e_f} B^{\downarrow \uparrow *}_{\vec e_i \vec e_f}=- B^{\downarrow \downarrow *}_{\vec e_i \vec e_f} B^{\uparrow \downarrow}_{\vec e_i \vec e_f}$. Using these properties together with the SU(2) invariance of $\opr H_{t-J}$, the RIXS intensity reduces to
\begin{align} \label{eq:rixs_fin}
    I_{RIXS}\left(\omega, k, \bar t\right)=& \frac{2}{L} \left( \abs{B^{\uparrow \uparrow}_{\vec e_i \vec e_f}}^2 + \abs{B^{\uparrow \downarrow}_{\vec e_i \vec e_f}}^2 \right) \int_{t_0}^\infty \matn d t'_1 \int_{t_0}^{\infty} \matn d t_1  s(t'_1,\bar t) s(t_1,\bar t) e^{-i \omega( t_1-t'_1)} \nonumber \\
    & \times \Bra{\Psi_0} \opr U_{s}(t_0,t'_1) \sum_{j'} e^{i k j'} p_{j' \uparrow}^\dagger \opr U_{t-J}(t'_1,t_1) p_{0 \uparrow} \opr U_{s}(t_1,t_0) \Ket{\Psi_0} 
\end{align}
with the according time evolution operators of $\opr H_{s}$ and $\opr H_{t-J}$. At this point we can recognize Eq.~(4) of the main text. The RIXS matrix elements in the ultra-short core-hole lifetime approximation can be calculated as
\begin{align}
    B^{\uparrow \uparrow}_{\vec e_i \vec e_f}= \frac{e^{*}_{f,z}}{\Gamma} \left( e_{i,y} - i e_{i,x} \right),\qquad B^{\uparrow \downarrow}_{\vec e_i \vec e_f}=\frac{e^{*}_{f,x}}{\Gamma} \left( i e_{i,x} - e_{i,y} \right).
\end{align}
The polarisation of the incoming and outgoing photons depends on the experimental setup, thus, we do not specify them here.

\subsubsection{RIXS vs ARPES}
Let us also compare the results obtained above to Angle-resolved photoemission spectroscopy (ARPES). ARPES is a photon-in electron-out technique that provides an experimental route for probing the single electron removal spectral function $A(k, \omega_i-\epsilon_f)$, where $\omega_i$ is the energy of the incoming photon and $\epsilon_f$ the energy of the photo-emitted electron \cite{Wang2018}. In most cases, ARPES is used to map $A(k, \omega_i-\epsilon_f)$ in the proximity of the Fermi level. Since it produces a charge excitation in the valence, this technique has been used to probe spin-charge separation in the quasi-one-dimensional compound SrCuO$_2$ \cite{Kim2006_NatPhys}. From the theoretical point of view one can describe the experimental spectrum with a $t-J$ model with parameters $J \sim 0.20$eV and $t \sim 0.60$eV (leading to a ratio $J/t \sim 0.33 < 2$) \cite{Kim1996}. In this case, the $t-J$ model allows for describing the dynamics of the holon in an antiferromagnetic background.

As explained in the previous section, RIXS is instead a photon-in photon-out technique that requires the excitation of a local core state, i.e. it requires the use of highly energetic photons. Thus, the final state in RIXS cannot possess any charge excitation since no electron leaves the sample during the process. However, one can still induce several other local or collective excitations, e.g. dd excitations, phonons or spin excitations. In the case of dd excitations this leads to the creation of an orbiton. As we showed in \secu{sec:t-Jmodel}, it is even possible to describe the dynamics of the orbiton in the antiferromagnetic background by an effective $t-J$ model ($t \sim 0.085$eV, $J \sim 0.238$eV, $J/t \sim 2.8 > 2$), and the corresponding equilibrium RIXS spectrum can be computed by considering the corresponding orbital spectral function $\chi(k, \omega_i-\epsilon_f)$ of the problem. 

The very different physical content of the two processes reflects on a different functional dependence of the effective $t$ and $J$ parameters by the original Hamiltonian interaction strength. 
%FG
In the previous sections of this work, 
%In our work, 
which is focused on the description of spin-orbital separation in Sr$_2$CuO$_3$, we report only the equations for the $t$ and $J$ parameters for the RIXS process, i.e. in the presence of an orbital excitation (see \eqref{eq:eff_mic_param_Undriv}). However, we can easily compute the hopping and superexchange expressions in the presence of a holon in Sr$_2$CuO$_3$, which would correspond to the ARPES process. In this case, $J$ is the same as we wrote in the \secu{sec:t-Jmodel}, while
\begin{align}
    t = \frac{\bar{t}_\sigma^2}{\Delta_x} \;,
\end{align}
with $\bar{t}_\sigma$ the renormalized hopping between the $3$d$_{x^2-y^2}$ orbital of a copper atom and the p$_x$ orbital of a nearest-neighbor oxygen, while $\Delta_x$ is the corresponding crystal field splitting between the states. By substituting the actual values for $\bar{t}_\sigma$ and $\Delta_x$, we obtain $t \sim 0.587$eV. The large difference in the values of the hoppings for ARPES and for RIXS comes from the fact that, in the former case, $t$ is second order in $\bar{t}_\sigma$, while in the latter, it is fourth order. Furthermore, we see that this leads to $J/t \sim 0.34 < 2$. In this case, the dynamics lie in the well-understood slow spinon regime.

\end{appendices}

%\bibliography{biblio}{}

%merlin.mbs apsrev4-1.bst 2010-07-25 4.21a (PWD, AO, DPC) hacked
%Control: key (0)
%Control: author (0) dotless jnrlst
%Control: editor formatted (1) identically to author
%Control: production of article title (0) allowed
%Control: page (1) range
%Control: year (0) verbatim
%Control: production of eprint (0) enabled
%

\end{document}